\documentstyle[preprint,aps,epsfig,tighten,floats]{revtex} 
\begin{document} 

\title{Superfluid Flow Past an Array of Scatterers}

\author{D. Taras-Semchuk} 
\address{Theory of Condensed Matter, Cavendish Laboratory, Cambridge
  CB3 OHE, United Kingdom} 

\author{J. M. F. Gunn}
\address{School of Physics and Astronomy, University of Birmingham,
Edgbaston, Birmingham B15 2TT, United Kingdom}

\maketitle 
 
\begin{abstract} 
We consider a model of nonlinear superfluid
flow past a periodic array of point-like scatterers in one dimension. 
An application
of this model is the determination of the critical current of a
Josephson array in a regime appropriate to a Ginzburg-Landau
formulation. Here, the array consists
of short normal-metal regions, in the presence of a
Hartree electron-electron interaction, and embedded within a
one-dimensional superconducting wire near its critical temperature,
$T_c$. We predict the critical current to depend linearly as
${\mathcal A} (T_c-T)$,
while the coefficient ${\mathcal A}$ 
depends sensitively on the sizes of the superconducting
and normal-metal regions and the strength and sign of the Hartree 
interaction. 
In the case of an attractive interaction, we find a further feature: 
the critical current vanishes linearly 
at some temperature $T^*$ less
than $T_c$, as well as at $T_c$ itself. We rule out a simple
explanation for the zero value of the critical current, at
this temperature $T^*$, in terms of order parameter fluctuations 
at low frequencies. 
\end{abstract} 


\section{Introduction}
The problem of nonlinear flow, as modeled by a
nonlinear Schrodinger equation in the presence of
a scattering potential,
relates to a variety of physical situations. 
One example is a weakly interacting Bose gas with impurities,
while another is a
Josephson array in a regime appropriate to a Ginzburg-Landau
formulation. Here, the array consists of 
short, normal-metal (N) regions in the presence of a
Hartree electron-electron interaction, and embedded within a
superconducting wire near its critical temperature, $T_c$. The latter
example falls within the rapidly developing field of study of the interplay of
the proximity effect and charging effects due to interactions
in disordered, inhomogeneous superconducting systems 
\cite{Sols,Zapata,Fazio,Nazarov,Huck}. 
Further
examples arise within nonlinear optics \cite{Talanov} and in the
study of gravity waves on deep water \cite{Hasimoto}. 

We will concern ourselves with the existence of time-{\em independent}
solutions of the flow equations in the presence of a supercurrent.
In general the value of the supercurrent is limited by a
maximum value, the critical current, 
above which the flow becomes unsteady and a time-dependent solution
must be sought. 
Recently Hakim \cite{Hakim} considered the superfluid flow 
past a single, repulsive scatterer in one dimension and
deduced the dependence of the critical current on the scattering
strength. We extend this model to the geometry of a regular
array of scatterers. We find a
rich dependence of the critical current
on the scatterer separation and strength, with
markedly different behaviour in the cases of 
repulsive and attractive scatterers.

Although we will focus here on the steady-state flow,
the physics of unsteady flows in related problems has also received attention.
For instance, unsteady flow past a single scatterer in one dimension 
\cite{Hakim} represents a transition state towards
the emission of solitons and hence phase-slip nucleation. 
It has also been pointed out \cite{Freire} that
steady-state flow 
may not be stable to mechanisms of quantum tunnelling,
leading to phase-slip nucleation and non-zero 
dissipation even below the (quasiclassical) critical current.
Higher dimensional analogues of the problem are 
also of theoretical and experimental importance, although
less tractable analytically:
recent work has simulated numerically
the flow of a superfluid past an obstacle \cite{Frisch} or through a
constriction \cite{Stone} in two dimensions, for which
vortices are nucleated above the critical supercurrent. These
investigations relate directly to experimental work \cite{Nancolas}
on superfluid $^4$He.

As mentioned above, an 
important application of this model is in the description of
supercurrent flow in a mesoscopic 
SNS device near the critical temperature of the S region.
The Ginzburg-Landau equations for the order parameter
 near the critical temperature of such a device
are equivalent to the flow equations
 for a superfluid in the Hartree approximation. Furthermore, a short N
 region may be modelled as a point-like scatterer whose strength
 diverges as $T_c$ is approached and the Ginzburg-Landau
 correlation length diverges.
We will exploit this equivalence to 
translate our results directly into experimental 
predictions for the
temperature dependence of the critical current of a Josephson array,
consisting of short N regions embedded in a superconducting wire near
$T_c$. 

The dependence of the critical current on the temperature is a
property that has already been examined in a number of related systems.
For example, it is an established result \cite{Tinkham}
that a superconducting wire (with no N
 regions) has a critical current that behaves as $(T_c-T)^{3/2}$ 
as $T_c$ is approached, while the introduction of
a single, short N region leads to a
 critical current that depends quadratically as $(T_c-T)^2$ \cite{Sols}.
In addition it has been possible to fabricate and measure experimentally
a superconducting `microladder' \cite{Pannetier},
consisting of a pair of S wires with connecting side-branches: 
in this geometry, a well-defined
correction to the $(T_c-T)^{3/2}$ dependance of the critical current of the
superconducting wire has been observed and
 explained theoretically. Courtois et al. \cite{Courtois}
have also measured the critical current for another
 geometry (somewhat complementary to that considered by us) of an array
 of short S regions near $T_c$ within a N metal, again with a
reasonable fit to the theory \cite{Fink}. 

The case of {\em two} short N regions, comprising a
double-barrier structure,
has also been examined theoretically, by Zapata and Sols \cite{Zapata}. 
They interpret such a system as
representing a nonlinear analogue of resonant tunnelling: such a
scattering structure is well known in a {\em linear} system to lead to
phenonema such as the sensitivity of the transmission
coefficient to the
scatterer separation, peaking near well-defined
resonances due to multiple inner reflection. 
The introduction of nonlinearity leads immediately to
markedly different behaviour, the added complexity precluding
even any kind of crossover regime in terms of a nonlinearity
parameter. The critical current of the double-barrier structure
depends as $(T_c'-T)^{1/2}$, where $T_c'$ is some temperature below
$T_c$.  

For the geometry of an array of short N
 regions embedded within a superconducting wire, we find 
a {\em linear} dependence of the critical current 
as ${\mathcal A}(T_c-T)$, in contrast to the cases described above. 
Furthermore,
the coefficient ${\mathcal A}$ depends sensitively 
on the sizes of the S and N regions and the strength and sign 
of the electron interaction.
The calculation of the chemical potential,
while trivial in the geometry of a finite number of scatterers, 
becomes more complex in the array geometry, and follows by 
the incorporation of a normalization condition to fix the total boson number. 
This procedure leads to the increased complexity of the flow solutions.

We find a further feature in the temperature dependence of the
critical current in the case of an attractive interaction: 
the critical current vanishes linearly 
at some temperature $T^*$ less
than $T_c$, as well as at $T_c$ itself. 
We will rule out a simple explanation for the
zero value of the critical current at this temperature, $T^*$,
in terms of order-parameter fluctuations at vanishingly low frequencies. 

The plan of this paper is as follows. In Section \ref{sec:model}, we
describe the model and flow equations for a superfluid in
the presence of scatterers. In Section
\ref{sec:single} we review briefly the solution of these equations for
a single scatterer in the time-independent regime. In Section
\ref{sec:array} we address the geometry of an array of scatterers and
present the results for the critical current as a function of
scatterer strength and separation. In Section \ref{sec:limit}
we derive analytically 
the form of the critical current in the various limiting cases of
scatterer strength and 
separation, while in Section \ref{sec:GL} we translate these
results into predictions for the temperature dependence of a Josephson
array near $T_c$. Section \ref{sec:sound} includes a discussion 
on order-parameter fluctuations at the special separation value of $L=2|g|$,
at which
the critical current vanishes, and Section \ref{sec:discuss} concludes 
with a summary.

\section{Model}
\label{sec:model}
We derive the relevant equations first for the problem 
of a one-dimensional superfluid in the presence of delta-function
scatterers. Taking the scatterers to have strength $g_{\alpha}$ 
and positions $r_{\alpha}$,
the Hamiltonian may be written
\begin{eqnarray}
H = \sum_{i}\left(-\frac{\hbar^2}{2m}\partial_{x_i}^2+\sum_{\alpha}g_{\alpha}
\delta(x_i-r_{\alpha})\right)
+\frac{1}{2}\lambda\sum_{i \neq k} \delta(x_i-x_k),
\label{Ham1}
\end{eqnarray}
where $i$ labels the $N$ bosons, which interact via 
a short-ranged, pairwise potential of strength $\lambda$.
We employ the Hartree approximation \cite{Nozieres} to write 
the ground state wavefunction in the symmetrised form,
\begin{eqnarray*}
\Psi(x_1,x_2,\ldots,x_N,t) = \frac{1}{{\mathcal
L}_{sys}^{N/2}} \prod_{i=1}^{N}\psi(x_i,t),
\end{eqnarray*}
where ${\mathcal L}_{sys}$ is the system size. 
This leads to the following nonlinear Schrodinger equation for 
$\psi(x,t)$:

\begin{eqnarray}
-\frac{1}{2}\nabla_x^2\psi+|\psi|^2\psi+\sum_{\alpha} g_{\alpha}
\delta(x-r_{\alpha}) \psi=i\partial_t\psi,
\label{nlse}
\end{eqnarray}
where we have rescaled length according to units of the healing
length, $\ell_h = \hbar/(\lambda n m)^{1/2}$, and energy (and the
$g_{\alpha}$) according to
units of the Hartree energy, $n\lambda$, where $n$ is the average
particle density. The speed of sound in the condensate in
the absence of the scatterers, ie. $(n\lambda/m)^{1/2}$, is equal to
unity with this choice of units.

In addition, we must enforce the condition of fixed total boson number,
which is achieved by normalizing the wavefunction according to
\begin{eqnarray}
\int |\psi|^2(x) dx = {\mathcal L}_{sys}.
\label{dfix1}
\end{eqnarray}

We now introduce the number-phase representation 
(the Madelung transformation \cite{Madelung}):
\begin{eqnarray}
\psi(x,t) = \sqrt{\rho(x,t)}\exp(i S(x,t)).
\label{n-phase}
\end{eqnarray}
Inserting eqn.~(\ref{n-phase})
into the Schrodinger equation, eqn.~(\ref{nlse}), 
gives the continuity equation and the
Bernouilli equation respectively,
\begin{eqnarray}
\partial_t \rho+\partial_x(\rho \partial_x S) &=& 0, 
\label{cont}\\
-\frac{1}{2}\frac{\partial_x^2\sqrt{\rho}}{\sqrt{\rho}}
+\partial_t S
+\frac{1}{2}(\partial_x S)^2 + \sum_{\alpha} g_{\alpha} \delta(x-r_{\alpha})
+\rho &=& 0,
\label{Euler}
\end{eqnarray}
while the normalization condition, (\ref{dfix1}), becomes
\begin{eqnarray}
\int \rho(x) dx = {\mathcal L}_{\rm sys}.
\label{dfix1b}
\end{eqnarray}
For time-independent flow, we may set
$\partial_t\rho$ to zero, while the phase $S(t)$ advances
uniformly in time according to the Josephson relation, $\partial_t S= -\mu$,
where $\mu$ is the chemical potential.
The continuity equation, eqn.~(\ref{cont}), then integrates to 
\begin{eqnarray}
\rho \partial_x S = j,
\label{1int}
\end{eqnarray}
where $j$, a constant, is the supercurrent density. 
Writing $\phi(x) = \rho(x)^{1/2}$, the flow equation,
eqn.~(\ref{Euler}), becomes
\begin{eqnarray}
-\frac{1}{2}\partial_x^2\phi+\phi^3-\mu\phi 
+\sum_{\alpha} g_{\alpha} \delta(x-r_{\alpha}) \phi
+\frac{j^2}{2\phi^3}= 0.
\label{GL}
\end{eqnarray} 
We will discuss the solution of this equation at length in the
following sections. We see that it is of a general
Ginzburg-Landau form, as we clarify further in section \ref{sec:GL}. 
We consider first the case of a single scatterer before turning to 
an array of scatterers.

\section{Single Scatterer}
\label{sec:single}

\begin{figure}[tbp]
\begin{center}
\epsfig{file=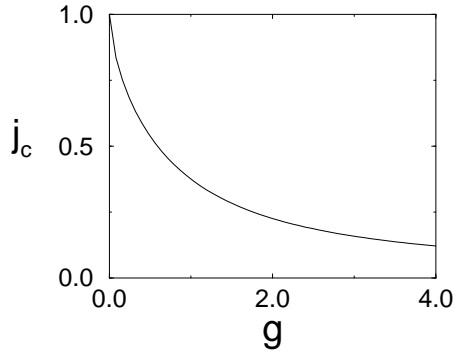,height=5cm}
\caption{\label{jcsinglefig}
Critical current, $j_c$, for a single, repulsive impurity of
  strength $g>0$.}
\end{center}
\end{figure}

In this section we review the results for the critical current of a
single scatterer \cite{Hakim}.
We take the scatterer to be
of strength $g$ and placed at a position $x=0$. The
flow eqn.~(\ref{GL}) gives the jump condition
\begin{eqnarray}
\frac{1}{2}\left[\partial_x\phi\right]_{0^-}^{0^+} = g \phi(0).
\label{jump1}
\end{eqnarray}
In addition, from the normalization
condition, eqn.~(\ref{dfix1b}), we enforce that
\begin{eqnarray}
\phi(x) \to 1 \quad {\rm as} \quad x\to \infty.
\label{infcond}
\end{eqnarray}
From eqn.~(\ref{infcond}) and the flow equation, eqn.~(\ref{GL}), we
find the chemical potential immediately as 
\begin{eqnarray*}
\mu = 1+\frac{j^2}{2}.
\end{eqnarray*}
With the above relations, the flow equation may be integrated as
follows:
\begin{eqnarray}
\begin{array}{rclr}
\rho(x) &=& j^2+(1-j^2)
\left\{\begin{array}{c} {\rm tanh}^2 \\ {\rm coth}^2
\end{array}\right\}
\left[\sqrt{1-j^2}|x| +\alpha\right],
& \left\{\begin{array}{c} g>0, \\ g<0. \end{array}\right.
\end{array}
\label{1imp}
\end{eqnarray}
The integration constant, $\alpha$, still needs to be determined from
the jump condition, (\ref{jump1}). For an attractive impurity, $g<0$,
the jump condition may be fulfilled for all values of the supercurrent
up to the speed of sound, $j\le1$. This is connected to the fact that
the value of $\rho$ at the impurity ($\rho(0)$)
may become arbitrarily large. 
Thus
the critical current for a single impurity, $j_{c,0}$, is in this case 
\begin{eqnarray*}
j_{c,0}(g<0) = 1.
\end{eqnarray*}
In contrast, for a repulsive impurity, $g>0$, the jump condition
can no longer be satisfied for all $j$ up to 1, due to the restriction
that $0\le \rho(0) \le 1$.
Instead we have that
$j_{c,0}(g>0) <1$. Although Hakim \cite{Hakim} 
obtained an implicit formula for the critical current,
we obtain in Appendix \ref{app:single}
a relatively simple explicit formula. The resulting dependence of
$j_{c,0}(g)$ on $g$ is illustrated in figure \ref{jcsinglefig}.
In the limits of
large and small scattering strength, $g$, we have  
\begin{eqnarray}
j_{c,0}(g>0) =\left\{\begin{array}{cc}
 1/(2g), & g \gg 1, \label{jbigg}\\
1-\frac{3}{4}(2g^2)^{1/3}, & g \ll 1.
\end{array}\right.
\end{eqnarray}

Note that the result for large $g$ will find relevance in section \ref{sec:GL}
when we examine the equivalent problem of 
the critical current in an SNS junction (with a single normal-metal
region) close to the critical temperature.
Before elaborating on this interpretation, we keep to the example of a
superfluid, and turn to the case of a periodic array of scatterers.

\section{Array of Scatterers}
\label{sec:array}
We now examine the case of a periodic array of scatterers, of equal
strength $g$ and separation $L$. We place them at positions $r_{\alpha} =
(2\alpha+1)L/2$, so that $\phi(x)$ is symmetric about $x=0$ and we may
restrict attention to $|x|<L/2$. We first show how the flow equations
may be integrated before discussing the results of their full
solution.

\subsection{Integration of the Flow Equations}
The
first integral of the flow equation, (\ref{GL}), 
may be written as follows:
\begin{eqnarray*}
(\partial_x\phi)^2 = \phi^4-2\mu\phi^2-\frac{j^2}{\phi^2} 
-\phi(0)^4+2\mu\phi(0)^2+\frac{j^2}{\phi(0)^2}.
\end{eqnarray*}
We may now factorize the right-hand side of this expression:
this operation greatly simplifies the following analysis. We find
\begin{eqnarray}
(\partial_x\phi)^2 =
\frac{1}{\phi^2}(\phi^2-\phi(0)^2)(\phi^2-\alpha)(\phi^2-\beta),
\label{firstint}
\end{eqnarray}
where
\begin{eqnarray}
\left\{\begin{array}{c}\alpha \\ \beta\end{array}\right\} =
\mu-\frac{\rho(0)}{2}\pm\left[\left(\mu-\frac{\rho(0)}{2}\right)^2
-\frac{j^2}{\rho(0)}\right]^{1/2}.
\label{abeta}
\end{eqnarray}
This leads to the following solution for $\rho(x)$:
\begin{eqnarray*}
\begin{array}{lr}\rho(x) = \beta+ (\rho(0)-\beta)\left\{\begin{array}{c}
{\rm sn}^2 \\ \frac{1}{k^2 {\rm sn}^2}\end{array}\right\}
(\sqrt{\alpha-\beta}\,x+{\rm Re} K(k),k), &
\left\{\begin{array}{c} g>0, \\ g<0,\end{array}\right.
\end{array}
\end{eqnarray*}
where the symmetry of $\rho(x)$ about the origin has been
automatically incorporated,
sn and $K$ are the elliptic integral and complete elliptic
functions respectively \cite{Gradsteyn}, and
\begin{eqnarray}
k = \sqrt{\frac{\rho(0)-\beta}{\alpha-\beta}}.
\label{keq}
\end{eqnarray}

It still remains to determine the  
two integration constants, $\mu$ and $\rho(0)$.
These are specified by the normalization condition corresponding to
eqn.~(\ref{dfix1b}), that is,
\begin{eqnarray}
\frac{2}{L}\int_0^{L/2}\rho(x)dx = 1,
\label{dfix2}
\end{eqnarray}
and the jump condition,
\begin{eqnarray}
\frac{1}{2}\left[\partial_x\phi
\right]_{(L/2)^-}^{(L/2)^+} = g \phi\left(\frac{L}{2}\right).
\label{jump2}
\end{eqnarray}
The latter condition may be rewritten, using the first integral in
eqn.~(\ref{firstint}), as follows:
\begin{eqnarray}
\frac{j^2}{\rho(0)\rho(L/2)} =
g^2\frac{\rho(L/2)}{\rho(L/2)-\rho(0)}+2\mu-\rho(L/2)-\rho(0).
\label{jump3}
\end{eqnarray}
This represents an implicit equation for $j$,
since the right-hand side is dependent on $j$ through $\mu$ and $\rho(0)$.
The problem of calculating the critical current of the array has now
been reduced to finding the maximum value of $j$ for which the
two conditions (\ref{dfix2}) and (\ref{jump3}) may be satisfied 
simultaneously. 

\subsection{Results}
\label{sec:results}
Having described the integration of the flow equations,
we present here the results for the critical current of the array.
Further details of the working towards these results will be described
in the following section.

It is clear that in the limit of $L\to
\infty$, $j_c$
must approach its single-impurity value, $j_{c,0}$:
\begin{eqnarray}
j_c \to j_{c,0}(g), \quad L \to \infty.
\label{bigL}
\end{eqnarray} 

\begin{figure}
\begin{center}
\[\begin{array}{ccc}
\epsfig{file=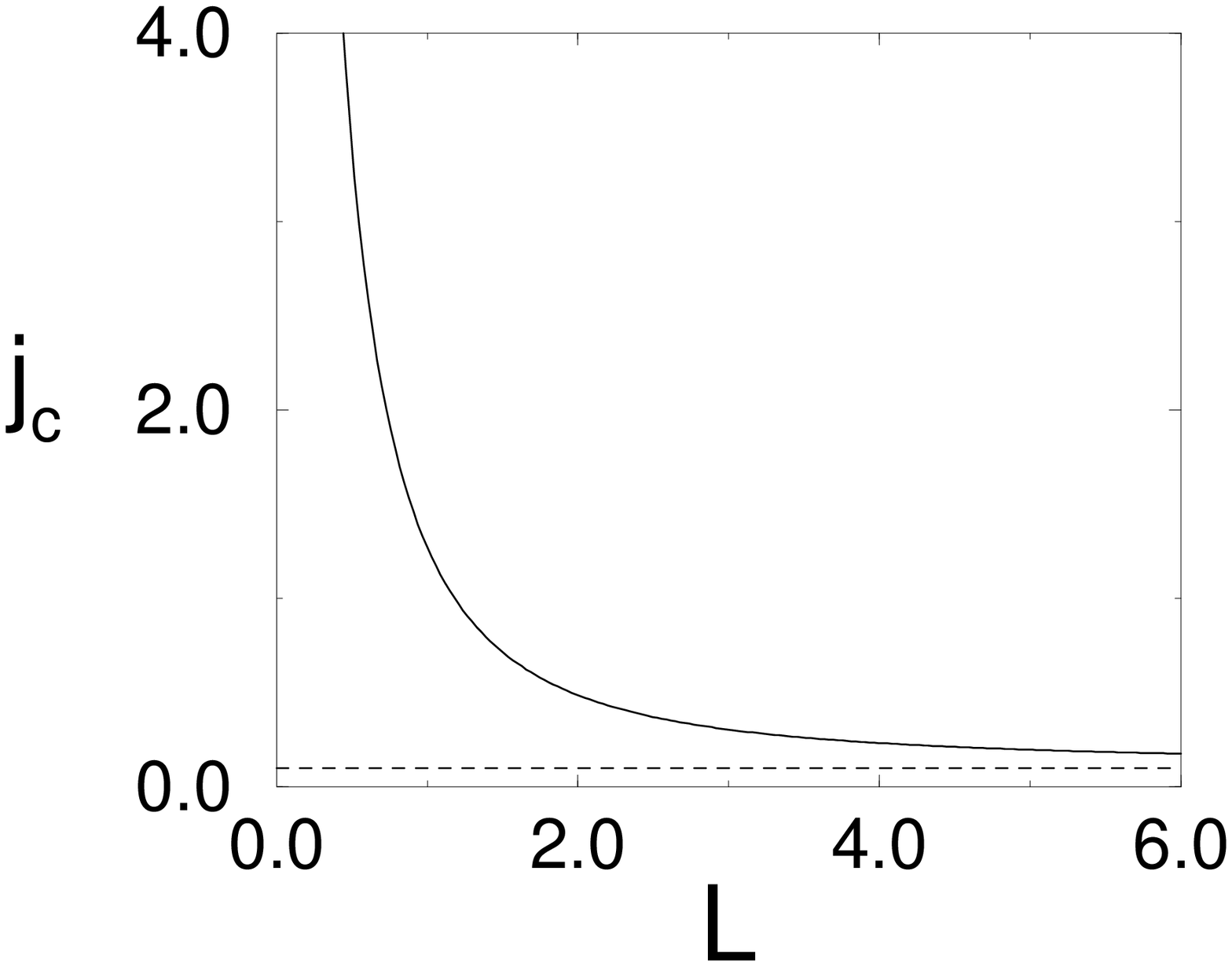,height=6cm} &&
\epsfig{file=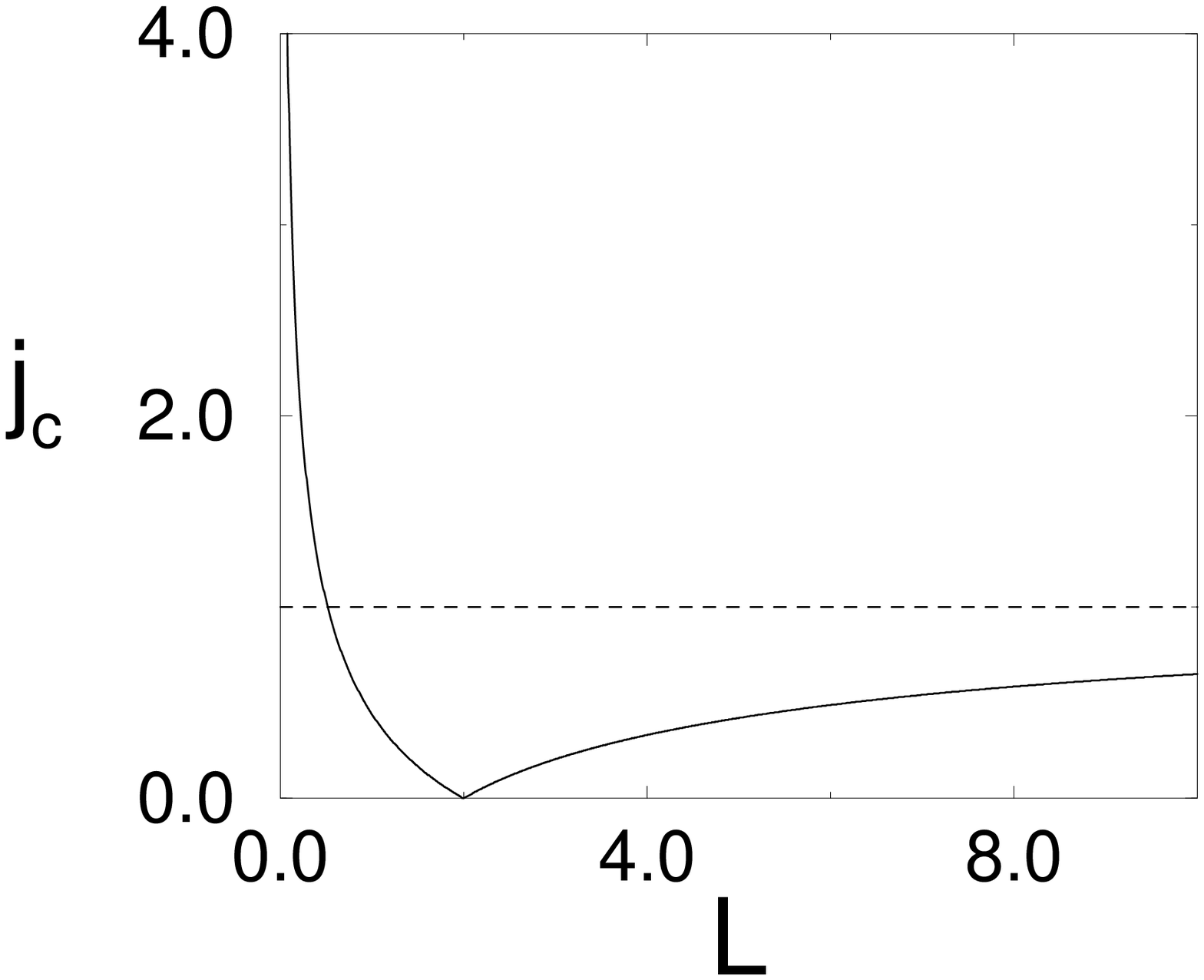,height=6cm} \\
{\em (a)} && {\em (b)} 
\end{array}\]
\caption{\label{fig:j1}Critical current, $j_c$ (solid line),
as a function of $L$ for 
{\em (a)} a repulsive array ($g=5$) and {\em (b)} 
an attractive array ($g=-1.0$). 
In each case, $j_c$ approaches the single-impurity value (the dotted line)
for large $L$.}
\end{center}
\end{figure}

For general values of the parameters $g$ and $L$, 
the determination of $j_c$ must
be performed numerically. Figure \ref{fig:j1} shows 
$j_c$ as a function of $L$ for a typical repulsive and attractive
case. We see that the correct behaviour is reproduced
in the limit of $L\to\infty$. We also see that in the attractive
case, the critical current vanishes altogether at one special value of
$L$. We will show below that this value is given by $L=2|g|$. We also
see that $j_c$ diverges at small $L$ in both the repulsive and
attractive case.

\begin{figure}
\begin{center}
\[\begin{array}{ccc}
\epsfig{file=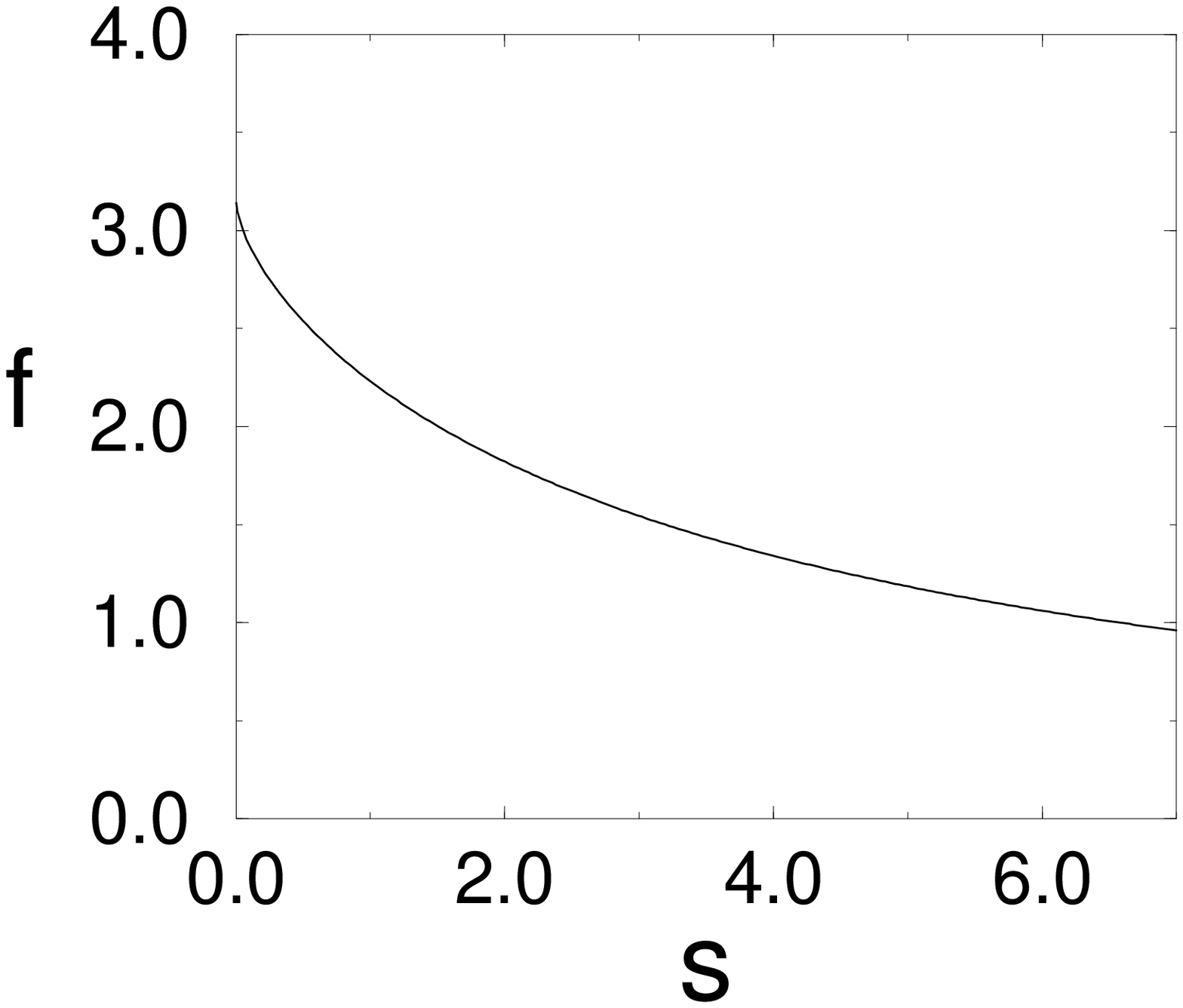,height=6cm} &&
\epsfig{file=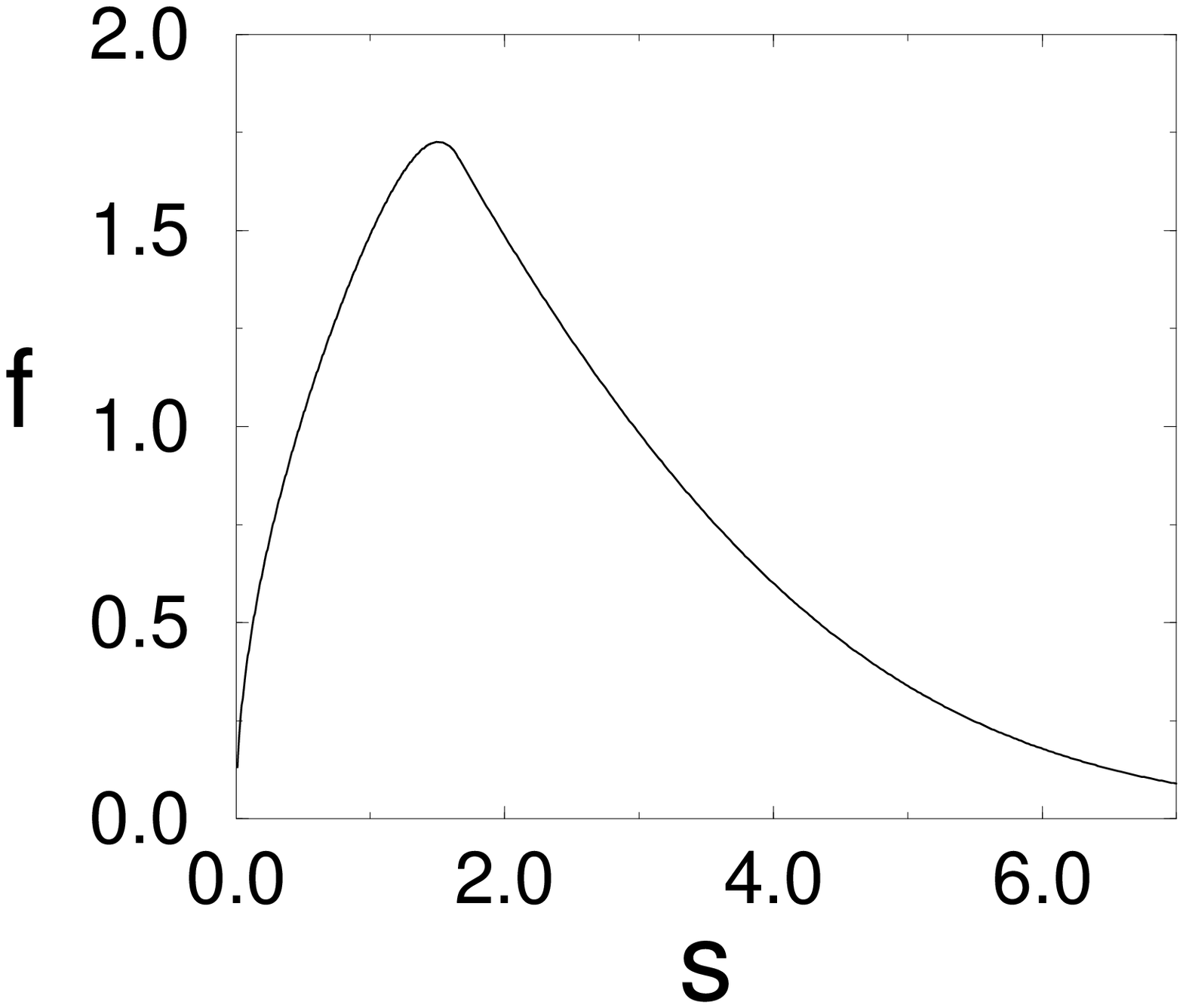,height=6cm} \\
{\em (a)} && {\em (b)} 
\end{array}\]
\caption{\label{fig:j2}
The scaling function $f$ as a function of $s$, for {\em (a)}
  the repulsive and {\em (b)} the attractive case. In the limit
  of $L\to 0$, the critical current is given by $j = f(gL)/L$.}
\end{center}
\end{figure}

In the limit of small scatterer separation, $L\to 0$, we find further that 
$j_c$ satisfies the following scaling form:
\begin{eqnarray}
j_c \to \frac{1}{L} f(g L), \qquad L\to 0.
\label{smallL}
\end{eqnarray}
The scaling function $f$ possesses a surprisingly nontrivial structure
which we have obtained 
numerically and show in figure \ref{fig:j2}. Notice the marked difference in
these forms for the repulsive and attractive cases. We see also that, 
with interpretation of this system as a superfluid as 
described in section \ref{sec:model}, the scaling
variable $|g|L$ is unaffected by the energy and length rescaling (in
units of the Hartree energy and healing length respectively) and
hence of the boson interaction $\lambda$.

The limits of large and small separation needs a more precise
definition, which in fact differs in the repulsive and attractive
cases: the limits correspond to $L \gg L_0$ and $L \ll L_0$ respectively,
where $L_0 = 1$ for the repulsive case and $L_0 = |g|$ for the
attractive case.

The scaling function $f$ will find direct relevance in the prediction
for the temperature-dependence of the critical current of a Josephson
array in section \ref{sec:GL}. In the following section, we will
demonstrate analytically the following limiting behaviour of this function:
\begin{eqnarray}
f(s, g>0) \simeq \left\{\begin{array}{cr}
\pi, & s \ll 1, \\
\frac{\pi^2}{s}, & s \gg 1,
\end{array}\right.
\label{frep}
\end{eqnarray}
and
\begin{eqnarray}
f(s, g<0) \simeq \left\{\begin{array}{cr}
\sqrt{2s}, & s \ll 1, \\
2s^2 e^{-s}, & s \gg 1.
\end{array}\right.
\label{fatt}
\end{eqnarray}
The critical current $j_c$ for small $L$ then follows by
$s=|g|L$ and the scaling relation (\ref{smallL}). 
We will also show analytically that the critical current
vanishes as the special point $L=2|g|$ is approached, 
in a linear fashion as $j_c \propto
|L/2|g|-1|$. 

\section{Limiting Forms of the Solutions}
\label{sec:limit}
In this section we derive analytically the limiting forms of the
scaling function given by eqns.~(\ref{frep}) and (\ref{fatt}), and
hence of the critical current at small separations. We take
the repulsive case first, which requires the limit $L \ll 1$, and then
the attractive case, which requires $L \ll |g|$. Note that the condition
itself for small separation is different in the repulsive and
attractive cases. While staying within this condition, we will examine
separately the subcases of $|g|L \ll 1$ and $|g|L \gg 1$.

In general, we find at such small separations the chemical potential
approaches a large positive (negative) value in the repulsive
(attractive) case, reflecting the large `potential energy' of the
scatterers. In addition, for small $|g|L$, the density $\rho(x)$ remains
close to 1 for all values of $x$. In contrast, for large $|g|L$,
the value of
$\rho(x)$ approaches 2 at the origin and 0 at the scatterers for the
repulsive case, and
0 at the origin and $|g|L\gg 1$ at the scatterers in the attractive case. 

To derive the critical current, it is necessary to identify the
somewhat subtle interplay of the various parameters of the problem,
which requires a separate and quite different discussion for each of
the four cases. We also include a discussion of the special points 
$L=2|g|$, at which the
critical current vanishes entirely. Our approach in each case
will be to indicate
the existence of a local (rather than global)
maximum in the supercurrent. Strictly
speaking, we still
need to justify these choices of a local maxima
as the relevant values for the critical current. To do so, we appeal
to the numerical results of section \ref{sec:results}
which establish the smooth interpolation from the limit of small
separation to large separation and hence to the single scatterer
result 
for the critical current, which certainly does represent a global maximum.
 
\subsection{Repulsive case at small separation ($L\ll 1$)}
In the repulsive case, the limit of small separation requires that
$L\ll 1$. We examine separately the subcases of $gL\ll 1$
and $gL \gg 1$, while staying within the limit of $L \ll 1$. 

In both subcases, we have that $\mu\gg 1$, and hence
$k\ll 1$ by eqn.~(\ref{keq}). The sn function is then well
approximated by a standard cosine function:
\begin{eqnarray}
\rho(x) =
\beta+(\rho(0)-\beta)\cos^2(\sqrt{2\mu}x),
\label{rform}
\end{eqnarray}
where $\beta = j^2/(2\mu\rho(0))$.
We also 
have that $\rho(x)$ must contain no more than one
half-oscillation between impurities, so that $\sqrt{2\mu}L/2 \in (0,\pi/2)$.
The normalization condition, eqn.~(\ref{dfix2}) consequently
simplifies to the form,
\begin{eqnarray}
1 = \beta+\frac{1}{2}(\rho(0)-\beta)
\left[1+\frac{1}{\sqrt{2\mu}L}\sin(\sqrt{2\mu}L)\right].
\label{dfix3}
\end{eqnarray}
As shown in Appendix \ref{app:rep}, this limiting form leads to
a critical current which behaves as $j_c \simeq \pi/L$ for
$gL \ll 1$ and $j_c \simeq \pi^2/(gL^2)$ for $gL \gg 1$ (while $L\ll 1$).

\subsection{Attractive case at small separation ($L\ll |g|$)}
In the attractive case, the limit of small separation requires that
$L \ll |g|$. Note that this condition itself is separate from that
in the repulsive case. Again, we will concern ourselves with the
two subcases $|g|L \ll 1$ and $|g|L \gg 1$, while staying within $L
\ll |g|$. In both subcases, the chemical potential is large and negative:
$|\mu| \gg 1$, $\mu<0$. 
The behaviour of $k$ however is different according to
the limit of $|g|L$: we have $k \gg 1$ for $|g|L \ll 1$ but $k\sim 1$ for
$|g|L \gg 1$. The sn function then reduces to either a cosine function
or a tanh function:
\begin{eqnarray}
\rho(x) &=& \left\{\begin{array}{cr}
\beta+(\rho(0)-\beta){\rm sec}^2(\sqrt{\rho(0)-\beta}x), & |g|L \ll 1,
\label{rforma1}
\\
\beta+(\rho(0)-\beta){\rm coth}^2(p-\sqrt{\alpha-\beta}x), & |g|L \gg 1,
\end{array}\right. 
\label{rforma2}\\
p &=& \frac{1}{2}\log\left(\frac{8}{k-1}\right).
\label{pdef}
\end{eqnarray}
Further details of the derivation of the critical current are
contained in Appendix \ref{app:att}: we find $j_c \simeq (2|g|/L)^{1/2}$
for $|g|L \ll 1$ and $j_c \simeq 2g^2L \exp(-|g|L)$ for $|g|L \gg 1$. This
concludes the derivation of the behaviour of the critical current and
hence the scaling function contained in eqns.~(\ref{frep}) and (\ref{fatt}).

\subsection{Attractive case in the limit $L\to 2|g|$}
In this section, we consider the attractive case at separations near
the special value,
$L=2|g|$, at which the critical current vanishes entirely. This limit is
characterized by a {\em divergence} in one of the parameters, namely 
$k\to \infty$. Consequently, $\mu\to\rho(0)/2$, and hence
\begin{eqnarray}
\rho(x) \to \rho(0) {\rm sec}^2(\sqrt{\rho(0)} x), \qquad L \to 2|g|.
\label{rformsp}
\end{eqnarray}
The normalization condition, eqn.~(\ref{dfix2}), and the jump
condition, eqn.~(\ref{jump3}), become equivalent in this limit, and
simplify to:
\begin{eqnarray}
\frac{L}{2} =
\sqrt{\rho(0)}\tan\left[\sqrt{\rho(0)}\frac{L}{2}\right].
\label{r0sln}
\end{eqnarray}
Eqn.~(\ref{r0sln}) for $\rho(0)$ is soluble for all values of $L$. 

At this point it is still not obvious that the special point $L=2|g|$
does not admit a steady solution at any non-zero
supercurrent. To demonstrate this fact, we consider the point $L=
2|g|(1+\epsilon)$ for some small $\epsilon$ (positive or negative),
and show that $j_c$ vanishes
as $|\epsilon| \to 0$. 

In this limit, we identify the small parameter 
$\gamma = \mu-\rho(0)/2$, $|\gamma| \ll 1$. We will find $\gamma$
to be proportional to $\epsilon$ at the critical current. Then 
\begin{eqnarray*}\left.
\begin{array}{c} \alpha \\ \beta \end{array}\right\}
 = \gamma \pm \left(\gamma^2-\frac{j^2}{\rho(0)}\right)^{1/2}.
\end{eqnarray*}
We find that $k^2= \rho(0)/(2(\gamma-\beta))$, and hence the limiting form
for the density is
\begin{eqnarray}
\rho(x) = \beta + (\rho(0)-\beta){\rm sec}^2\left[\sqrt{\rho(0)-\beta}
\left(1-\frac{(\gamma-\beta)}{2\rho(0)}\right)x\right].
\label{denss}
\end{eqnarray}
As we show in Appendix \ref{app:spec}, this limiting form leads to a
critical current $j_c$ that vanishes linearly in $|\epsilon|$ as claimed.

In section \ref{sec:sound} 
we will examine order parameter fluctuations, propagating as sound
waves, at this separation
$L=2|g|$. We will rule out the existence of such fluctuations at arbitrarily
low frequencies as a simple explanation for the
zero value of the critical current at this value of the separation. 
Instead we may view this effect as a consequence of
the nonlinearity of the flow equations. 

\section{Josephson Array near $T_c$}
\label{sec:GL}
\begin{figure}
\begin{center}
\epsfig{file=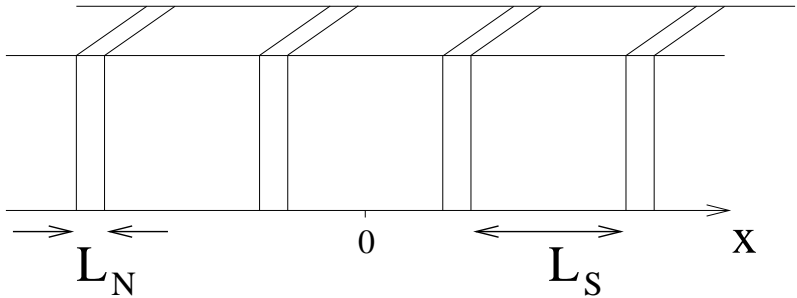,height=4cm}
\caption{\label{fig:array}The geometry of the Josephson array.}
\end{center}
\end{figure}

Having the determined the behaviour of the critical current for the
periodic array of scatterers, 
we now show how these results may be translated directly in
experimental predictions for the temperature dependence of the
critical current of a Josephson
array in a regime where a Ginzburg-Landau formulation is appropriate.
Here, the array is near the critical temperature, $T_c$, of its S
regions, and has
the quasi-1D geometry shown in
figure \ref{fig:array}: the S regions are of length $L_S$ while the N
regions are of length $L_N \ll L_S$ and subject to a Hartree potential, $V$.

We first show that the Ginzburg-Landau 
equations for the array are of the same form as the flow equation,
eqn.~(\ref{GL}), for a superfluid. Furthermore, as shown by Zapata et
al. \cite{Zapata}, the short normal-metal
regions,
in the presence of the Hartee potential,
may be mimicked by a series of repulsive or attractive
delta-function scatterers, depending on the sign of the interaction $V$. 

The Ginzburg-Landau equations for the order parameter $\Psi(x)$
of the S region may be written as \cite{L+L}
\begin{eqnarray*}
-\frac{\hbar^2}{4m}\partial^2_x \Psi +a(T)\Psi+b|\Psi|^2\Psi = 0,
\end{eqnarray*}
where $a(T)$ is negative and proportional to $(T_c-T)$, 
$b = 2\pi\kappa^2e^2\hbar^2/m^2c^2$ and $\kappa$ is the
Ginzburg-Landau parameter. Writing $\Psi = |\psi|\exp(iS)$ and defining
the supercurrent (per unit of cross-sectional area) 
as $j = (e\hbar/m)|\psi|^2\partial_x S$, we have
\begin{eqnarray}
-\frac{\hbar^2}{4m}\partial^2_x \psi +a(T)\psi +b\psi^3
+\frac{mj^2}{4e^2\psi^3} = 0.
\label{GL2}
\end{eqnarray}
In addition to the above differential equation for $\psi$, we need to
specify appropriate boundary conditions. In the {\em absence} of any N
regions, the order parameter of the homogeneous superconductor, and
hence condensate density, is fixed at the value $\psi_S(T)$. 
Furthermore, for a geometry containing only a 
{\em single} N region, 
the appropriate boundary condition is to set that $\psi(x)$
approaches $\psi_S$ for limiting values of $x$ far from the N
region, by analogy with the boundary condition used in section
\ref{sec:single} in considering a single scatterer. For an array,
however, no
such condition at infinity can apply; instead, we fix the total
boson number at a certain temperature to equal that in the absence of
the N regions. This gives the condition
\begin{eqnarray*}
\frac{1}{L_S} \int_{-L_S/2}^{L_S/2} \psi(x) dx = \psi_S,
\end{eqnarray*}
in similarity to the normalization condition, eqn.~(\ref{dfix2}), of
section \ref{sec:array}.
Here the integration extends over a single S region.
To proceed, we rescale the position variable and current as follows:
\begin{eqnarray}
y &=& \frac{x}{\sqrt{2}\xi(T)}, 
\label{yT}\\
j &=& \frac{\sqrt{2}\hbar c^2}{16\pi\kappa^2e\xi(T)^3} J,
\label{jT}
\end{eqnarray}
where the coherence length is defined as
\begin{eqnarray*}
\xi(T) &=& \frac{\hbar}{2(b m)^{1/2}\psi_S} \\
&=& \xi_0'\left(1-\frac{T}{T_c}\right)^{-1/2},
\end{eqnarray*}
and $\xi_0' = 0.74\xi_0$ for a clean superconductor and $\xi_0' = 0.85
\sqrt{\ell \xi_0}$ for a dirty superconductor, where $\xi_0$ and
$\ell$ are the zero-temperature coherence length and mean free path
respectively.
In these rescaled variables, the Ginzburg-Landau equation, eqn.~(\ref{GL2}),
becomes
\begin{eqnarray}
-\frac{1}{2}\partial^2_y \phi -\mu\phi+\phi^3+\frac{J^2}{2\phi^3} = 0,
\label{GL3}
\end{eqnarray}
where $\mu = |a|/(b \psi_{\infty}^2)$. We see that for the geometry of
a single N junction, the chemical potential becomes $\mu =
1+J^2/2$. We also see that eqn.~(\ref{GL3})
reproduces the flow equation (\ref{GL}) as expected, together with the
normalization condition (\ref{dfix2}).

Notice that the rescaling leads immediately, by eqn.~(\ref{jT}),
to a $(T_c-T)^{3/2}$
dependence for the critical current of a S wire with no N regions, in
accordance with established theory (see e.g. \cite{Tinkham}).
If we now include one or more short N regions,
the flow equation in an N region
in the rescaled variables, corresponding to eqn.~(\ref{GL3}), reads as
\begin{eqnarray}
-\frac{1}{2}\partial_y^2\phi +\theta \phi =-\frac{J^2}{2\phi^3},
\label{NGL}
\end{eqnarray}
where 
\begin{eqnarray*}
\theta &=& V/(b\psi_S^2) \\
&=& {\rm sgn}(V) \frac{\xi(T)^2}{\xi_N^2}, 
\end{eqnarray*}
and $\xi_N^2 = \hbar^2/(4m|V|)$ is the 
Ginzburg-Landau correlation length in the normal-metal. 
Following Ref.~\cite{Zapata}, we may now integrate trivially the
flow equation (\ref{NGL}) over the whole N region (located in $y \in
(y^-,y^+)$, say) to give
\begin{eqnarray}
-\frac{1}{2}\left[\partial_y \phi\right]_{y^-}^{y^+} =
  \frac{L_N}{\sqrt{2}\xi(T)}\theta \phi,
\label{jumpGL}
\end{eqnarray}
as long as
\begin{eqnarray*}
L_N \ll \ \xi_N.
\end{eqnarray*}
Notice that we have
dropped the current-dependent term on the right-hand side of
eqn.~(\ref{NGL}) as is consistent for a sufficiently short N region
near $T_c$ (see \cite{Zapata}). 
By identification of eqn.~(\ref{jumpGL})
with eqn.~(\ref{jump2}), we see that we
may model the short N regions by delta-function
scatterers of strength
\begin{eqnarray}
g = {\rm sgn}(V) \frac{L_N \xi(T)}{\sqrt{2}\xi_N^2}.
\label{gdef}
\end{eqnarray}
At the same time, following the rescaling of the
size of the S regions, $L_S$, under the
transformation (\ref{yT}), we may also identify the scatterer
separation as
\begin{eqnarray}
L = \frac{L_S}{\sqrt{2}\xi(T)}.
\label{Ldef}
\end{eqnarray}
By eqns.~(\ref{gdef}) and (\ref{Ldef}),
we see that the limit of $T\to T_c$, and hence $\xi(T) \to\infty$,
corresponds to taking $g\to\infty$ and $L\to 0$ simultaneously, 
such that the product $gL (= L_N L_S/(2\xi_N^2))$ is fixed.

Given the above information we are now in a position to predict the
temperature dependence of the critical current of the array. 
For orientation however we start with
the simpler geometry of a {\em single} S-N-S junction with a repulsive
interaction. As the temperature
approaches $T_c$, $g$ diverges by eqn.~(\ref{gdef})
and we insert the large $g$ limit of the critical current for a
single impurity, eqn.~(\ref{jbigg}): $J_c = 1/(2g)$. Relation 
(\ref{jT}) then gives the critical current of the SNS junction as
\begin{eqnarray*}
j_c(T) =  \frac{\hbar c^2\xi_N^2}{16\pi e \kappa^2 (\xi_0')^2 L_N}
\left(1-\frac{T}{T_c}\right)^2
\end{eqnarray*} 
 per unit area, ie. it varies {\em quadratically} as $(T_c-T)^2$.

We turn now to the  array. As the temperature approaches $T_c$,
we have $L\to 0$ and we become able to apply the scaling form
(\ref{smallL}) for the critical current per unit area:
\begin{eqnarray*}
j_c(T) &=& \frac{\hbar c^2}{8\pi e \kappa^2 \xi_0'} 
f(s) \left(1-\frac{T}{T_c}\right), \\
s &=& \frac{L_N L_S}{2\xi_N^2}.
\end{eqnarray*}
Recall that 
$f(s)$ is the scaling function discussed in section \ref{sec:array},
illustrated in figure \ref{fig:j2} for the repulsive and attractive
cases and obeying the limiting forms contained in eqns.~(\ref{frep})
and (\ref{fatt}). 
We see that we have a {\em linear}
dependence of the critical current as $j_c = {\mathcal A}(T_c-T)$. 
In addition, the
associated coefficient, ${\mathcal A}$, of this linear dependence 
depends on the parameters $L_N$, $L_S$ and $V$ through the 
product $L_N L_S V$:
\begin{eqnarray*}
{\mathcal A} = \frac{\hbar c^2}{8\pi e\kappa^2\xi_0' T_c} f\left(
\frac{2m L_N L_S V}{\hbar^2}\right).
\end{eqnarray*}
Notice that both large and small values of this product
may be probed while staying within the requirement that $L_N \ll
\xi_N$.

In moving away from $T_c$ for the attractive ($V<0$) case, an addition feature
arises in accordance with the vanishing of the critical current at
$L=2|g|$: we have that the critical current vanishes at the temperature
$T^*<T_c$, as well as at $T_c$ itself. Here, $T^*$ is determined by
\begin{eqnarray*}
\xi(T^*) = \sqrt{\frac{L_S}{2L_N}}\xi_N.
\end{eqnarray*}
Moreover, since $j_c$ vanishes linearly in $|L/2|g|-1|$, we have 
that the critical current vanishes linearly in $|T-T^*|$ as $T^*$
is approached.

This concludes our discussion of the temperature dependence of the
critical current of the Josephson array near $T_c$ (and $T^*$). 
We now focus on the case of $T\to T^*$ to examine whether 
order parameter fluctuations 
exist at arbitrarily low frequencies in this limit.

\section{Order Parameter Fluctuations at $L=2|g|$}
\label{sec:sound}
Having identified the special value ($L=2|g|$) of the (attractive) 
scatterer separation
at which the critical current takes a zero value, we
here examine whether a simple explanation for this effect exists in
terms of order parameter fluctuations at arbitrarily low frequencies. 

We use a standard procedure to describe such fluctuations in a 
condensate of non-uniform density
(see, for example, Giorgini et al.~\cite{Giorgini}): we perturb the
wavefunction $\psi(x,t)$ (see Section \ref{sec:model})
by a small contribution that is oscillatory in time:
\begin{eqnarray*}
\psi(x,t) = e^{-i\mu t}\left[\phi(x) +u(x)e^{-i\omega t}+v^*(x)
e^{i\omega t}\right].
\end{eqnarray*}
The flow equation (\ref{nlse}) may then be linearised in these small
oscillations to give the following coupled flow equations for $u(x)$
and $v(x)$:
\begin{eqnarray}
{\mathcal L}u(x)+\rho(x)v(x) &=& \omega u(x), 
\label{sound1}\\
\rho(x) u(x) + {\mathcal L}v(x) &=& -\omega v(x),
\label{sound2}
\end{eqnarray}
where 
\begin{eqnarray*}
{\mathcal L} = -\partial_x^2/2 -\mu+2\rho(x)+\sum_{\alpha} g_{\alpha}
\delta(x-x_{\alpha}).
\end{eqnarray*}
These equations are supplemented with the normalization condition,
\begin{eqnarray}
\int dx \left[ u^*(x)u(x)-v^*(x)v(x)\right] = 1,
\label{norms}
\end{eqnarray}
while trivial integration of the flow equations (\ref{sound1}) and
(\ref{sound2}) over each scatterer gives the jump conditions,
\begin{eqnarray}
\frac{1}{2}\left[\partial_x \left(\begin{array}{c} u \\ v
\end{array}\right)\right]_{x_{\alpha}^-}^{x_{\alpha}^+} 
= g \left(\begin{array}{c} u \\ v
\end{array}\right)(x_{\alpha}).
\label{jumpu}
\end{eqnarray}
Our aim is to determine whether fluctuations in the wavefunction,
propagating as sound waves, exist at arbitrarily low 
frequencies at the separation $L=2|g|$. To this end, we 
search for a consistent solution for $u(x)$ and $v(x)$ in the limit of
$\omega \to 0$.

For preparation we review the solutions in the homogeneous case, ie. in
the absence of any scatterers. In this case, the perturbations are
plane-waves,  $u(x) =
\exp(-ikx)u/{\mathcal L}_{\rm sys}$ and $v(x) = \exp(-ikx)v/{\mathcal
  L}_{\rm sys}$, where ${\mathcal L}_{\rm sys}$ is the system size as
before. The flow equations (\ref{sound1}) and (\ref{sound2}), together
with the normalization condition, (\ref{norms}), are now easily solved to give
\begin{eqnarray}
u^2 &=& \frac{{\mathcal L}+\omega}{2\omega}, \label{usol}\\
v^2 &=& \frac{{\mathcal L}-\omega}{2\omega}, \label{vsol}
\end{eqnarray}
where ${\mathcal L} = k^2/2+\rho$. The dispersion relation
reads
\begin{eqnarray*}
\omega^2 = \frac{1}{2}k^2(k^2+2\rho).
\end{eqnarray*}

We now generalise to the case of the non-uniform density at the
separation, $L=2|g|$. In this case, we need to incorporate the jump
condition given by (\ref{jumpu}) at each scatterer. Given the
periodicity of the arrangement of the scatterers, we will search for
purely periodic solutions for $u$ and $v$:
\begin{eqnarray}
\left(\begin{array}{c} u \\ v \end{array}\right)(x+L) = \pm 
\left(\begin{array}{c} u \\ v\end{array}\right)(x),
\label{bloch}
\end{eqnarray}
so that we may restrict attention to only a single region $|x| < L/2$
with one jump condition. Eqn.~(\ref{bloch}) may be seen as a Bloch theorem
for $u$ and $v$, at Bloch 
momenta which are precisely multiples of $\pi/L$: 
odd and even multiples give rise to solutions in $u$ and $v$ which are
odd and even in $x$,  respectively.

Bearing in mind the form of the
solutions (\ref{usol}) and (\ref{vsol}) for the uniform case, a
consistent series expansion for 
$u(x)$ and $v(x)$ in the limit of $\omega \to 0$ becomes:
\begin{eqnarray*}
u(x) &=& \frac{u_0(x)}{\sqrt{\omega}}(1+{\mathcal O}(\omega)), \\
v(x) &=& \frac{v_0(x)}{\sqrt{\omega}}(1+{\mathcal O}(\omega)).
\end{eqnarray*}
The normalization condition, (\ref{norms}), to zeroth order
then gives $u_0^2=v_0^2$:
by analogy with the homogeneous case, we take $u_0= v_0$. The zeroth-order
differential equation for $u_0(x)$, corresponding to
eqn.~(\ref{sound1}), becomes
\begin{eqnarray*}
 \left[-\frac{1}{2}\partial_x^2-\mu+3\rho(x)\right]u(x) = 0,
\end{eqnarray*}
for $x \in (-L/2,L/2)$. 
We substitute the form (\ref{rformsp}) for the density
$\rho(x)$: we find
\begin{eqnarray}
\left[\partial_y^2+(1-6\,{\rm sec}^2y)\right] u(y) &=& 0, 
\label{diffu}
\end{eqnarray}
where $y = \sqrt{\rho(0)}x$.
The two independent solutions of this differential equation
are as follows:
\begin{eqnarray*}
u_1(y) = \frac{\sin y}{\cos^2 y}; \qquad u_2(y) =
F\left(1,2;\frac{7}{2};\cos^2y\right)\cos^3y,
\end{eqnarray*}
where $F$ is Gauss's hypergeometric function \cite{Gradsteyn}.
By the Bloch theorem, eqn.~(\ref{bloch}), we will take $u(y)$ to 
be either purely odd or purely even: 
$u(y) = A u_1(y)$ or $u(y) = B u_2(y)$. We 
now check whether either solution is
compatible with the jump condition. We see that this check is
independent of the coefficients $A$ and $B$. Taking the odd solution first,
$u_1(y)$, the jump condition (\ref{jumpu}) becomes
\begin{eqnarray*}
\frac{L}{2} 
&=& \sqrt{\rho(0)}\left[{\rm cot}(y_0)+2\tan(y_0)\right],
\end{eqnarray*}
where $y_0 = \sqrt{\rho(0)}L/2$. This simplifies by use of
eqn.~(\ref{r0sln}) to
\begin{eqnarray*}
0 = \rho(0) + \frac{L^2}{4},
\end{eqnarray*}
which clearly cannot be satisfied for any L, as the right-hand side
always exceeds zero. This solution must therefore be discarded.
We are left with the even solution, $u_2(y)$, for which the
jump condition (\ref{jumpu}) becomes
\begin{eqnarray*}
\frac{L}{2} =
\sqrt{\rho(0)}\left[-\frac{5\cos^2(y_0)}{\sin(y_0)u(y_0)}+
{\rm cot}(y_0)+2\tan(y_0)\right],
\end{eqnarray*}
which follows from standard properties of the hypergeometric function
\cite{Gradsteyn}. Use of eqn.~(\ref{r0sln}) simplifies this condition
to the form
\begin{eqnarray}
F\left(1,2;\frac{7}{2};\cos^2(y_0)\right) = 5.
\label{cont2}
\end{eqnarray}
However it is easy to check that $0 < F(1,2;7/2;z) < 5$
for $0<z<1$, with $F(1,2;7/2;0) = 0$ and $F(1,2;7/2;1) = 5$: clearly
the condition (\ref{cont2}) cannot be satisfied either 
for any $L>0$. 

We conclude that a periodic solution for $u(x)$ does not exist in the limit of
$\omega \to 0$ and hence such order parameter fluctuations 
do {\em not} exist at arbitrarily
low frequencies. Instead it seems we must
appeal to the nonlinearity of the flow
equations as an underlying cause for the surprising effect of a zero
critical current at $L=2|g|$.

\section{Summary}
\label{sec:discuss}
In this paper we have examined the superfluid flow past an array of point-like
scatterers in one dimension. We have determined the critical current
of the flow, above which the flow becomes unsteady. While the
result for a single scatterer is recovered in the limit of large
scatterer separation, we find a scaling form for the critical current
in the opposite limit of small scatterer separation. 
The scaling function takes a
particular form, separate in the repulsive and attractive cases, that
we have obtained numerically, as well as 
derived analytically in the various limiting cases of
scatterer strength and separation. We also find the additional feature
in the attractive case that the critical current vanishes altogether at
one special value of the scatterer separation ($L=2|g|$).

While these results are applicable to a variety of physical
situations, an important application is in the prediction of the
temperature dependence of a Josepshon array, in the presence of a
Hartee potential and near $T_c$. In contrast to
dependencies already derived and observed experimentally in other
geometries, we find for the array
a linear dependence of the critical current as
${\mathcal A}(T_c-T)$. 
The coefficient ${\mathcal A}$ depends sensitively on the size of the
normal regions ($L_N$) and of the superconducting regions ($L_S$)
and the Hartree interaction, $V$, through the product $L_S L_N
V$. In addition, for the attractive case ($V<0$),
the critical current is suppressed to zero as $T\to
T^*$, as well as at $T_c$ itself, 
where $T^*$ is some temperature less than $T_c$. We have
ruled out a simple explanation for this suppression of the critical
current in terms of order parameter fluctuations at low frequencies, instead
appealing to the nonlinearity of the flow equations as an underlying cause.

\section*{Acknowledgements}  
One of us (DT-S) gratefully acknowledges the financial support of
EPSRC and Trinity College.

\appendix
\section{Single impurity: critical current}
\label{app:single}
To find the critical current for a single impurity,
we replace the integration constant $\alpha$ in eqn.~(\ref{1imp})
in favour of the constant
$\rho(0)$, where $\tanh^2\alpha = (\rho(0)-j^2)/(1-j^2)$, and
reformulate the jump condition (\ref{jump1}) as
\begin{eqnarray}
g^2 \rho(0)^2 = (\rho(0)-j^2)(\rho(0)-1)^2,
\label{j4}
\end{eqnarray}
where we have used the first integral of the flow equation, eqn.~(\ref{GL}).
The critical current, $j_{c,0}$, is determined by the condition $\partial
j_c/\partial \rho(0)=0$, or
\begin{eqnarray}
2g^2\rho(0) = (\rho(0)-1)(3\rho(0)-1-2j_{c,0}^2).
\label{max}
\end{eqnarray}
Eliminating $g$ from eqns.~(\ref{j4}) and (\ref{max}), we find
\begin{eqnarray}
\rho(0) = \frac{1}{2} \left[-1+(1+8j_{c,0}^2)^{1/2}\right],
\label{eq1}
\end{eqnarray}
while eliminating $j$ in a similar manner gives
\begin{eqnarray}
(\rho(0)-1)^3+2g^2\rho(0) = 0.
\label{eq2}
\end{eqnarray}
Eqns.~(\ref{eq1}) and (\ref{eq2}) give the following solution for $j_{c,0}$:
\begin{eqnarray*}
j_{c,0}(g>0)^2 &=& 1-\frac{2g^2}{3}+\frac{3g}{\sqrt{2}}\left(-\frac{R}{3}
+\frac{1}{R}\right)+g^2\left(\frac{R^2}{9}+\frac{1}{R^2}\right), \\
R^3&=& \frac{2\sqrt{2}g}{-1+\sqrt{1+8g^2/27}}.
\end{eqnarray*}
This formula reduces to the forms shown in the main text in the limits
of large and small scattering strength, $g$.


\section{Repulsive impurities at Small Separation} 
\label{app:rep}
In this section we 
obtain the critical current for repulsive impurities at small
separation, $L\ll 1$. We treat the subcases of $gL \gg 1$ and $gL \ll 1$
separately.

\subsection{The subcase $gL \gg 1$}
In this limit, the have that the density $\rho(x)$ approaches 2 at the
origin and zero at the impurities. Using eqn.~(\ref{rform}),
we see that the latter limit means that 
$\sqrt{2\mu}L/2 \simeq \pi/2$. Writing
\begin{eqnarray}
\sqrt{2\mu}\frac{L}{2} = \frac{\pi}{2}-\gamma,
\label{gammadef}
\end{eqnarray}
where $\gamma \ll 1$, we have from eqn.~(\ref{rform}) that
\begin{eqnarray}
\rho(L/2) \simeq  \frac{j^2L^2}{2\pi^2}+2\gamma^2.
\label{rL2}
\end{eqnarray}
In addition, the normalization condition, eqn.~(\ref{dfix3}), leads to
$\rho(0) = 2(1-2\gamma/\pi)$.
Inserting eqn.~(\ref{rL2}) into the jump condition, eqn.~(\ref{jump3}), we find
\begin{eqnarray*}
j^2 &\simeq& -g^2\rho(L/2)^2+4\mu\rho(L/2) \\ 
&=& -g^2(\frac{j^2L^2}{2\pi^2}+2\gamma)^2
+j^2+\frac{4\pi^2\gamma^2}{L^2}
\end{eqnarray*}
or
\begin{eqnarray*}
j =
\frac{\pi^2}{gL^2}\left[1-\left(1-\frac{2Lg\gamma}{\pi}\right)^2\right]
^{1/2}.
\end{eqnarray*}
Maximisation of this expression with respect to $\gamma$
is now trivial 
and gives
\begin{eqnarray*}
j_c = \frac{\pi^2}{gL^2}, \quad gL \gg 1,
\end{eqnarray*}
as required. At this critical current, we have $\gamma=\pi/(2gL)\ll 1$ and
$\rho(L/2) = \pi^2/(gL)^2\ll 1$.

\subsection{The subcase $gL \ll 1$}
We now take the opposite limit of $gL \ll 1$. In this case, we have
that the density $\rho(x)$ is close to 1 for all values of $x$.
We set $u = \rho(0) - 1 \ll 1$,
and define $\gamma$ as before, in eqn.~(\ref{gammadef}). As may be
verified at the end of the calculation, we have that $u,\gamma \ll
1$ (for $j$ near its critical value). 
We first use the normalization condition, eqn.~(\ref{dfix3}),
and the jump condition, eqn.~(\ref{jump3}), to determine a relation
between $u$ and $\gamma$. This will allow us to obtain an expression
for $j$ in terms of only $\gamma$, which may then be maximised simply.

The normalization condition, eqn.~(\ref{dfix3}), now reads
\begin{eqnarray*}
1 = \beta+\frac{\rho(0)-\beta}{2}\left(1+\frac{2\gamma}{\pi-2\gamma}\right),
\end{eqnarray*}
leading to 
\begin{eqnarray}
\beta = 1-u+\ldots
\label{betaeq1}
\end{eqnarray}
Since $\beta = j^2/(2\mu\rho(0))$, this gives
\begin{eqnarray}
\frac{j^2L^2}{\pi^2} = 1-\frac{4\gamma}{\pi}-u^2 +\ldots
\label{jexp}
\end{eqnarray}
In eqn.~(\ref{betaeq1}) and (\ref{jexp}) we have kept to first order
in $\gamma$ and second order in $u$, which may be checked to be an
appropriate level of accuracy at the end of the calculation.
The limiting form for the density, eqn.~(\ref{rform}) now gives
\begin{eqnarray*}
\rho(L/2) &=& \beta+(\rho(0)-\beta)\gamma^2 \\
&=& \frac{j^2}{2\mu\rho(0)}+2u\gamma^2.
\end{eqnarray*}
Inserting this into the jump condition, (\ref{jump3}), we find
\begin{eqnarray*}
j^2 &=& -\frac{g^2}{2u}+j^2+4\mu u\gamma^2.
\end{eqnarray*}
Notice the cancellation of the $j^2$ terms, which allows us to find
\begin{eqnarray}
u\gamma = \frac{g L}{2\pi}.
\label{ugam}
\end{eqnarray}
Inserting eqn.~(\ref{ugam}) into eqn.~(\ref{jexp}), we find
\begin{eqnarray*}
\frac{j^2L^2}{\pi^2} =
1-\frac{4\gamma}{\pi}-\frac{(gL)^2}{4\pi^2\gamma^2}
+\ldots
\end{eqnarray*}
This expression may now be maximised simply with respect to
$\gamma$, with the result
\begin{eqnarray}
j_c =
\frac{\pi}{L}\left[1-\frac{3}{2}\left(\frac{gL}{\pi^2}\right)^{2/3}\right],
\label{jexp2}
\end{eqnarray}
as required. At the critical current, we have that
\begin{eqnarray*}
\gamma = \frac{1}{2\pi^{1/3}}(gL)^{2/3} \ll 1; \quad 
u =\frac{1}{\pi^{2/3}}(gL)^{1/3} \ll 1.
\end{eqnarray*}

\section{Attractive impurities at Small Separation}
\label{app:att}
In this section we 
obtain the critical current for attractive impurities at small
separation, $L\ll |g|$. We again treat the subcases of $|g|L \gg 1$ 
and $|g|L \ll 1$ separately.

\subsection{The subcase $|g|L \gg 1$}
In the limit of $|g|L \gg 1$, we have that $\rho(x)$ is close to zero at
the origin and $\rho(L/2) \gg 1$ at the impurities. As may be confirmed at
the end of the calculation, we have $|\beta|
\gg |\alpha|$ and hence $k\simeq 1$, or more precisely, 
\begin{eqnarray}
k = 1+\frac{\rho(0)+|\alpha|}{4|\mu|},
\label{ke2}
\end{eqnarray}
where $\alpha = -j^2/(2\mu\rho(0))$.
The jump condition (\ref{jump3}) gives
\begin{eqnarray*}
0 \simeq g^2 -2|\mu|,
\end{eqnarray*}
and hence $\rho(x) = g^2 {\rm csch}^2(p-|g||x|)$. The normalization
condition, eqn.~(\ref{dfix2}), then leads to
\begin{eqnarray*}
1 = \frac{4|g|}{L}  \exp(-2p+|g|L),
\end{eqnarray*}
and hence $\rho(L/2) = |g|L \gg 1$. The expression for $k$, eqn.~(\ref{ke2}),
together with the definition of $p$, eqn.~(\ref{pdef}), together give
\begin{eqnarray*}
j^2 = g^2\rho(0)(4|g|L \exp(-|g|L)-\rho(0)),
\end{eqnarray*}
which is again easily maximised with respect to the parameter
$\rho(0)$, to give $j_c = 2g^2L\exp(-|g|L)$ as required. At the
critical current, we have that $\rho(0) = 2|g|L\exp(-|g|L) \ll 1$, $\alpha =
-2|g|L\exp(-|g|L) \ll 1$, and $|\beta| = 2g^2 \gg 1$.

\subsection{The subcase $|g|L \ll 1$}
In the limit of $|g|L \ll 1$, 
we have that $\rho(x)$ is again close to 1 for all values
of $x$. Consider the limiting expressions, eqn.~(\ref{abeta}), for the
parameters $\alpha$ and $\beta$: we see that taking $\alpha \to \beta
\simeq \mu-\rho(0)/2$
and hence
$j^2 \to (\mu-\rho/2)^2\rho(0)$ leads to a local maximum for the supercurrent
 as the solutions loses its applicability above such values of $j$. In
 this limit, we have $k \gg 1$ and the form (\ref{rforma1}) follows
 for the density. The normalization condition, (\ref{dfix2}), gives
\begin{eqnarray}
\left(|\mu|+1+\frac{1}{2}\rho(0)\right)\frac{L}{2} =
\left(|\mu|+\frac{3}{2}\rho(0)\right) \frac{L}{2},
\end{eqnarray}
where we have used  
$|\mu|L^2 \ll 1$, which we will find to hold given $|g|L\ll 1$. We
see that the normalization condition is satisfied automatically as
$\rho(0) \simeq 1$. We also have that $\rho(L/2) \simeq
\rho(0)+j^2L^2/4$, and it may be checked retrospectively 
that $\rho(L/2)\simeq 1$ for
$|g|L \ll 1$. The jump condition, (\ref{jump3}), then leads to
\begin{eqnarray*}
j_c^2 = \frac{4g^2}{j_c^2L^2}-2j_c,
\end{eqnarray*}
and hence $j_c \simeq (2|g|/L)^{1/2}$ as required.

\section{Attractive Impurities at $L\to 2|g|$}.
\label{app:spec}
In this section, we consider the special value of the separation,
$L=2|g|$, at which the critical current vanishes entirely in the
attractive case. More specifically, we take $L= 2|g|(1+\epsilon)$ and
consider the limit of very small $\epsilon$ (which may be positive or 
negative). Using the limiting form
of the density, eqn.~(\ref{denss}, in the main text, we see that the
boson-number fixing condition, eqn.~(\ref{dfix2}), leads to 
\begin{eqnarray*}
\rho(L/2) = \rho(0) +
(1-\beta)^2\frac{L^2}{4}\left(1-\frac{\gamma-\beta}{\rho(0)}\right).
\end{eqnarray*}
Inserting this value for $\rho(L/2)$ into the jump condition,
eqn.~(\ref{jump3}), we find
\begin{eqnarray*}
\frac{j^2}{\rho(0)\rho(L/2)^2} =
-\epsilon+\beta\left[2-\frac{1}{\rho(0)}\right]+
\gamma\left[\frac{1}{\rho(0)}+\frac{2}{\rho(L/2)}\right].
\end{eqnarray*}
At this point we are justified in substituting the values of $\rho(0)$
and $\rho(L/2)$ by their values at $\gamma=0$: that is, we 
take $\rho(0)$ as the solution
of eqn.~(\ref{r0sln}), while $\rho(L/2) = \rho(0)+L^2/4$.
This leads to the following expression for $j^2$:

\begin{eqnarray*}
j^2 = -\frac{\rho(0)}{(2\rho(0)-1)^2}\left[\epsilon\rho(0)-\left(
1+\frac{2\rho(0)}{\rho(L/2)}\right)\gamma\right]
\left[\epsilon\rho(0)-\left(4\rho(0)-1+2\frac{\rho(0)}{\rho(L/2)}\right)
\gamma\right].
\end{eqnarray*}
It may be verified that for $\epsilon=0$, ie. $L=2|g|$, the solution
$j=0$ exists for $\gamma=0$, while as soon as $\gamma$ becomes
non-zero, the solution for $j$ becomes imaginary: the flow equations
are unable to admit a well-behaved solution for any
non-zero supercurrent. In contrast, for non-zero
$\epsilon$, we may maximise $j$ straightforwardly
with respect to $\gamma$ to arrive at
\begin{eqnarray}
j_c^2 = \frac{\epsilon^2\rho(0)^3}
{(2\rho(0)/\rho(L/2)+1)(4\rho(0)+2\rho(0)/\rho(L/2)-1)}.
\end{eqnarray}
In particular, we see that $j_c$ vanishes linearly as $|\epsilon|$ as
$\epsilon \to 0$.
\end{document}